\documentclass[12pt]{article}

\usepackage{tikz}
\usepackage{epsfig,latexsym,amsfonts,amsmath,amsthm,amssymb,amsbsy,multirow,slashed,wasysym,textcomp,subfigure,wrapfig,datetime,comment,mathtools,cancel,cite,twistor,mathrsfs}%,showlabels}
\usepackage[hidelinks]{hyperref}
\usepackage[font={footnotesize},bf]{caption}

\def\ee{\end{equation}}
\def\be{\begin{equation}}
\def\bea{\begin{eqnarray}}
\def\eea{\end{eqnarray}}
\newcommand{\beq}{\begin{eqnarray}}
\newcommand{\eqq}{\end{eqnarray}}
 \newcommand{\badat}{\begin{alignedat}}
 \newcommand{\eadat}{\end{alignedat}}

\newcommand{\eal}[1]{\be \begin{aligned} #1 \end{aligned}\end{equation}} 
\newcommand{\eqn}[1]{\be #1 \end{equation}} 
\newcommand{\eqa}[1]{\bea  #1\end{eqnarray}}

\renewcommand{\d}{\mathrm{d}}

\long\def\new#1\endnew{{\bf #1}}		
\long\def\del#1\enddel{}

\def\del{\partial}

\def\re{\mathrm{e}}
\usepackage{color}

\newcommand{\pink}[1]{\textcolor{\pink}{#1}}

\definecolor{dblue}{rgb}{0.2,0.50,0.80}

\newcommand{\zb}{\bar{z}}

\def\O{\mathcal{O}}

\newcommand{\hb}{\bar{h}}

\newcommand{\bpd}{\bar{\partial}}

\def\bz{{\bar z}}

\def\s{ {\sigma} }

\renewcommand{\Im}{\mathrm{Im}}

\numberwithin{equation}{section} % equation numbers follow sections

\begin{document}
\begin{titlepage}
  \thispagestyle{empty}
  \begin{flushright}
  %\today
    \end{flushright}
  \bigskip
  \begin{center}
	 \vskip2cm
  \baselineskip=13pt {\LARGE \scshape{%Non-distributional 3-Gluon\\
  \vspace{0.5em} A Celestial Dual for MHV  Amplitudes }}

	 \vskip2cm
   \centerline{Walker Melton, Atul Sharma, Andrew Strominger and Tianli Wang}
 \vskip.5cm
 \noindent{\em Center for the Fundamental Laws of Nature,}
  \vskip.1cm
\noindent{\em  Harvard University,}
{\em Cambridge, MA, USA}
\bigskip
  \vskip1cm
  \end{center}
  \begin{abstract}
  
%It is shown  that a 2D CFT consisting of a central charge $c$ Liouville theory, a chiral level one, rank $N$ Kac-Moody algebra, and a weight $-3/2$ free fermion holographically generate 4D MHV tree-level flat space scattering amplitudes. The correlators of this 2D CFT give directly the 4D  leaf amplitudes associated to a single hyperbolic slice of flat space. The 4D amplitudes arise  in a large-$N$ and semiclassical large-$c$ limit, according to the holographic dictionary, as the translationally-invariant pole in the net conformal weight of the scattering states. A step in the demonstration is showing that the semiclassical limit of Liouville correlators are given by contact  AdS$_3$ Witten diagrams.   

It is shown  that a 2D CFT consisting of a central charge $c$ Liouville theory, a chiral level one, rank $N$ Kac-Moody algebra and a weight $-3/2$ free fermion
holographically generate 4D MHV tree-level scattering amplitudes. The correlators of this 2D CFT give directly the 4D  leaf amplitudes associated to a single hyperbolic slice of flat space. The 4D celestial amplitudes arise in a large-$N$ and semiclassical large-$c$ limit, according to the holographic dictionary, as a translationally-invariant combination of leaf amplitudes. A step in the demonstration is showing that the semiclassical limit of Liouville correlators are given by contact  AdS$_3$ Witten diagrams.

  \end{abstract}
% \noindent
%	

%
\end{titlepage}
\tableofcontents

\section{Introduction}

A central goal in celestial holography is to construct simple toy models in which 4D scattering amplitudes are holographically realized by 2D CFT correlators. An obstacle to this endeavor has been that the translation invariance of 4D amplitudes of massless particles implies that the low-point correlators are distributional, which is not the case for most familiar 2D CFTs. This obstacle has been circumvented by considering a variety of 4D contexts in which translation (but not Lorentz) invariance is broken \cite{Costello:2022wso, Fan:2022elem, Casali:2022fro, Gonzo:2022tjm, Stieberger:2023fju, Costello:2022jpg, Stieberger:2022zyk,Melton:2022def, Bittleston:2023bzp, Costello:2023hmi, Adamo:2023zeh, Ball:2023ukj, Melton:2023} or by shadow or light transforming the external particle states \cite{Sharma:2021gcz, Jorge-Diaz:2022dmy,Fan:2021isc,Fan:2021pbp,Chang:2022jut, Chang:2022seh}. But a direct holographic reconstruction of translationally-invariant 4D scattering amplitudes has so far not been obtained. 
  
In this paper we obtain  such  a direct reconstruction of 4D MHV gluon amplitudes from a 2D CFT. Our work relies heavily on a recent refined  analysis of the dictionary for flat space holography  including a complete accounting of causal light-cone singularities \cite{Melton:2023bjw, Melton:2024jyq}. It was shown that the full 4D amplitudes can be expressed as integrals over leaf amplitudes associated to the AdS$_3$ leaves of a hyperbolic foliation of flat space.  Each leaf amplitude has  bulk and boundary representations which are exactly those of familiar AdS holography, and the associated ``leaf CFTs'' have the corresponding familiar 2D singularity structure. They may be regarded as the primary building blocks of celestial holography. A simple formula was derived for the full celestial amplitudes as a combination of leaf amplitudes. At three points, celestial amplitudes are extracted from leaf amplitudes as a pole in the net conformal weight of the external particles. The distributional form of low-point celestial amplitudes, as mandated by translation invariance, then arises from cancellations between subamplitudes. We refer the reader to \cite{Melton:2023hiq,Melton:2023bjw,Melton:2024jyq} for  details.
  
In this work, we consider a non-unitary 2D CFT consisting of a Liouville field $\phi$ coupled to $N$ weight $(h,\hb) = ({1 \over 2},0)$ real free fermions $\psi_j$ plus a a single  weight $(-{3 \over 2},0)$ free fermion $\eta$. The $N$ fermions lead to a level one  $\SO(N)$ Kac-Moody current $J^a$.\footnote{We may similarly obtain an $\SU(N)$ gauge group by using  complex fermions.} Positive helicity $\SO(N)$ gluons are identified with Kac-Moody currents dressed by Liouville fields. Negative helicity gluons are then obtained by a further dressing with an $\eta$ bilinear. These are shown to match, in the appropriate limit, the MHV leaf amplitudes computed in \cite{Melton:2023bjw}. The leaf-to-celestial dictionary \cite{Melton:2023bjw} then precisely reproduces the known MHV scattering amplitudes, including the Parke-Taylor factor, the momentum-conserving delta function, and the various $\Theta$ functions separating different relative causal configurations of the asymptotic gluons. A key step along the way, detailed in appendix \ref{app:liouville}, is the demonstration that semiclassical Liouville correlators are given by scalar contact AdS$_3$ Witten diagrams. 

Duality here is demonstrated to leading (non-trivial) order expansions about certain limits of both the bulk and the boundary. 
First, from the bulk point of view, local conformal invariance is equivalent to the subleading soft graviton theorem \cite{Cachazo:2014fwa} which of course holds only in a theory of gravity. Hence we do not ever expect an exact celestial CFT dual to a gauge theory.  However, if we take a large-$N$ limit of gravity coupled to a gauge theory with a gauge group of rank $N$, gravity is suppressed. This suggests that leading order large-$N$ gauge theories may have celestial duals which are limits of 2D CFTs.
Accordingly we herein match only the leading-$N$ correlators of the 2D Kac-Moody  currents. Furthermore, we consider only tree-level MHV amplitudes in the bulk, which are matched to the large-$c$  semiclassical limit of the boundary Liouville theory.\footnote{ In AdS$_3$ holography, finite $c$ corrections correspond to loop corrections in the bulk theory. It is tempting to speculate that finite-$c$ corrections may play a similar role here, but it is not obvious why the 2D theory given here should be the full story beyond the MHV sector. Non-MHV amplitudes arise even at tree level in the bulk, so to get all the amplitudes will require some modification of the 2D CFT.}
  
This work does {\it not} comprise the sought-after full holographic duality between a 2D CFT and a 4D quantum theory of gravity. Rather we have achieved  the more limited goal of finding a duality between  limits  of a subset of a 2D boundary theory with those of a 4D bulk theory. 
 
We wish to stress that this paper is an amalgam of several important precursors.  The MHV leaf amplitudes appear in a slightly different context in \cite{Casali:2022fro}.  A very similar fermion system including the $\eta$ field is in \cite{Bu:2022dis} (itself deriving from  the prescient work \cite{Nair:1988bq})  and our construction was partly inspired by those in twisted holography \cite{Costello:2022wso, Costello:2022jpg}. An important connection to light operators in the semi-classical Liouville was noted  in a slightly different context in \cite{Stieberger:2022zyk, Taylor:2023bzj, Stieberger:2023fju}.  Finally, the connection between 2D Liouville and 4D gauge theory found by AGT \cite{Alday:2009aq} may be related. 

%%%%%%%%%%%%%%%%%%%%%%%%%%%%%%%%
%%%%%%%%%%%%%%%%%%%%%%%%%%%%%%%%

\section{The Celestial CFT}

In this section we describe  a relatively simple 2D CFT that generates 4D MHV amplitudes. This CFT has three components: the classical ($b \to 0$) limit of Liouville theory, a set of $N$ free fermions $\psi^i$ and a free chiral fermion $\eta$ of weight $-3/2$.
% \textcolor{blue}{with action
% \begin{equation}
%     S = \int d^2z [\rho\bar{\partial}\eta+\psi^i\bar{\partial}\psi^i + \frac{1}{4\pi}\bar{\partial}\phi\partial\phi+ \mu e^{2b\phi}].
% \end{equation}
We will refer to this as the dressed Liouville theory. In the next section, we describe the holographic dictionary through which MHV amplitudes are reproduced in the classical ($b \to 0$, $N\to\infty$) limit of dressed Liouville theory.

%%%%%%%%%%%%%%%%%%%%%%%%%%%%%%%%
%%%%%%%%%%%%%%%%%%%%%%%%%%%%%%%%

\subsection{Chiral fermion sector} 

The free fermion operators have the OPEs, for $i, j=1,\dots,N$, 
\begin{equation}
\begin{split}
    \psi^i(z)\,\psi^j(w) &= \frac{\delta^{ij}}{z-w}+ :\!\psi^i\psi^j\!:\!(z)+ O(z-w)\\
    \eta(z)\,\eta(w) &=(z-w)\,\eta \partial\eta(z) + O((z-w)^2)
\end{split}
\end{equation}
where $\psi^i$ have weights $(1/2,0)$ and $\eta$ has weight $(-3/2,0)$.  %\footnote{As the $\eta\eta$ OPE is nonsingular, we drop the normal order sign for operators built out of $\eta$.}

The  $\psi^i$ generate a level-1 Kac-Moody current algebra for SO($N$) with currents
\begin{equation}
    J^a(z) = \frac{1}{2}\,T^a_{ij}:\!\psi^i\psi^j\!:\!(z)
\end{equation}
where $T^a_{ij}$ are generators of $\mathfrak{so}(N)$ in the fundamental representation. We will also employ the dimension $-1$ chiral operators\footnote{We follow here the notation of \cite{Bu:2022dis} in which a related operator appears.}
\begin{equation}
    \bar J^a(z) = \eta \p \eta J^a(z)\,,
\end{equation}
with normal ordering being implicit. With this free field realization, one finds the OPEs
 \begin{align}
J^a(z)\,J^b(w) &\sim \frac{\delta^{ab}}{(z-w)^2}+ {\im f^{ab}{}_cJ^c(w)\over z-w} + O((z-w)^0)\,,\nonumber\\
J^a(z)\,\bar J^b(w) &\sim   \frac{\delta^{ab}\eta\partial\eta(w)}{(z-w)^2} +{\im f^{ab}{}_c\bar J^c(w)\over z-w} + O((z-w)^0),\\
\bar J^a(z)\,\bar J^b(w) &\sim {(z-w)^2\delta^{ab}\over 12}\,\eta\partial\eta\partial^2\eta\partial^3\eta(w) + \frac{\im f^{ab}{}_c(z-w)^3}{12}\,\eta\partial\eta\partial^2\eta\partial^3\eta J^c(w) + O((z-w)^4)\,.\nonumber
\end{align}
$J^a$ and $\bar J^a$ will enter the operators dual to positive and negative helicity gluons respectively.

Since $\eta$ has weight $h=-3/2$, it has four zero modes on the sphere. Correlation functions of $J^a$ and $\bar J^a$ must have exactly 2 $\bar J^a$ insertions to soak up these zero modes \cite{Bu:2022dis}. 
% The two-point function $\langle\bar{J}\bar{J}\rangle=\langle\eta\partial\eta\eta\partial\eta\rangle\langle JJ\rangle$ is computed as follows.
The relevant correlation function $\langle\eta\partial\eta(z_1)\,\eta\partial\eta(z_2)\rangle$ is computed as follows. The globally holomorphic zero modes of $\eta$ are
\begin{equation} 
\eta=\eta_{3/2}+\eta_{1/2}z+\eta_{-1/2}z^2 + \eta_{-3/2}z^3\,.
\end{equation}
These are the only modes contributing to the correlator, and lead to
\begin{equation} 
\langle\eta\partial\eta(z_1)\,\eta\partial\eta(z_2)\rangle
= z_{12}^4
\end{equation}
where we adopted the convention $\int\d\eta_{n}\,\eta_{n}=1$ for the Grassmann measure.

The $n$-point nonvanishing $J^a,\bar J^a$ correlators are 
\begin{equation}
    \big\langle\bar J^{a_1}(z_1)\,\bar J^{a_2}(z_2)\prod_{j=3}^n J^{a_j}(z_j) \big\rangle = \mathrm{Tr}\,(T^{a_1}T^{a_2}\cdots T^{a_n})\,\frac{z_{12}^4}{z_{12}z_{23}\cdots z_{n1}} + \cdots
\end{equation}
where $\cdots$ includes other color orderings and multi-trace terms. At leading order   in the large $N$ limit only the first term contributes to the color-ordered correlator which we denote with suppressed gauge  indices
\begin{equation}
    \big\langle\bar J(z_1)\,\bar J(z_2)\prod_{j=3}^n J(z_j) \big\rangle = \frac{z_{12}^4}{z_{12}z_{23}\cdots z_{n1}}\,.
\end{equation}

%%%%%%%%%%%%%%%%%%%%%%%%%%%%%%%%
%%%%%%%%%%%%%%%%%%%%%%%%%%%%%%%%

\subsection{Liouville sector}
We consider Liouville theory with central charge
\begin{equation}
    c_L = 1+6Q^2 = 1+6\,(b+b^{-1})^2\,, 
\end{equation}  and coupling $\mu$. This theory contains the primary operators  
\begin{equation}
    V_\alpha(z,\zb) = \re^{2\alpha\phi(z,\zb)}
\end{equation}
where $\phi$ is the Liouville field. They are said to have ``momentum''  $\alpha$ and have conformal weight
\begin{equation}
    \Delta(V_\alpha) = 2\alpha(Q-\alpha)\,.
\end{equation}
A brief review of the relevant aspects of Liouville  is provided in Appendix \ref{app:liouville}.

We are interested in correlation functions of light operators in the classical limit of Liouville theory. These operators have momenta $\al=b\s$ scaling  as $b$ in the $b \to 0$ limit. We show in  Appendix \ref{app:liouville} that these are proportional to $n$-point contact Witten diagrams in hyperbolic 3-space $H^3$,
\begin{equation}\label{dsx}
    \begin{split}
        \Big\langle\prod_{j=1}^n V_{b\s_j}(z_j,\zb_j)\Big\rangle &= \frac{\re^{-2\gamma_E+2/b^2}\lambda^{1/b^2}}{\pi b^3}\csc\big(\pi(b^{-2}-\tfrac{1}{2}\,\beta)\big)\,\lambda^{-1-\frac\beta2}\;\cC_{2\s_1,\ldots,2\s_n}
        \end{split}\,.
\end{equation}
Here, $\gamma_E$ is the Euler-Mascheroni constant, $\lambda = \pi \mu b^2$ is held fixed as we send $b\to0$, and $\beta$ is given by the sum over weights 
\be 
\beta = 2\,\big({\textstyle\sum_j\s_j}-2\big)\,,
\ee
and the contact Witten diagrams are given by
\be \label{lff} \cC_{2\s_1,\ldots,2\s_n}(z_i,\bz_i) = \int_{H^3}\mathrm{D}^3x\, \prod_{j=1}^n G_{2\s_j}(z_j,\bz_j;x)\,,
\ee
with $x$ denoting coordinates and $\mathrm{D}^3x$ the measure on the unit hyperboloid. The integrand is a product of scalar bulk-to-boundary propagators $G_{2\s_j}(z_j,\bz_j;x)$ of weight $2\s_j$.\footnote{ This  relation between semiclassical Liouville  and Witten diagrams appears to be new: related formulae 
appear in  \cite{Harlow:2011ny,Seiberg:1990eb}.} In the special case of $n=3$, we have the simple formula 
\be \cC_{2\s_1,2\s_2,2\s_3}=\frac{\pi}{2}\,\frac{\Gamma(\s_1+\s_2+\s_3-1)\,\Gamma(\s_1+\s_2-\s_3)\,\Gamma(\s_2+\s_3-\s_1)\,\Gamma(\s_3+\s_1-\s_2)}{\Gamma(2\s_1)\,\Gamma(2\s_2)\,\Gamma(2\s_3)\,(z_{12}\zb_{12})^{\s_1+\s_2-\s_3}(z_{23}\zb_{23})^{\s_2+\s_3-\s_1}(z_{31}\zb_{31})^{\s_3+\s_1-\s_2}}\,.
\ee

%%%%%%%%%%%%%%%%%%%%%%%%%%%%%%%%
%%%%%%%%%%%%%%%%%%%%%%%%%%%%%%%%

\section{Celestial Amplitudes from Dressed Liouville Correlators}

In this section, we explain the holographic dictionary from the 2D dressed Liouville theory to the 4D MHV amplitude. From this point on, we simply follow the prescription detailed in \cite{Melton:2023hiq,Melton:2023bjw,Melton:2024jyq} for the construction of celestial amplitudes from leaf amplitudes. We sketch it here but refer the reader to those references for details.

%%%%%%%%%%%%%%%%%%%%%%%%%%%%%%%%
%%%%%%%%%%%%%%%%%%%%%%%%%%%%%%%%

\subsection{Dressed Liouville $\to$ Euclidean leaves}

The first step is to state the dictionary between bulk conformal primary gluons and boundary dressed Liouville operators. 
We posit\footnote{Of course there are other operators in the dressed Liouville theory which are not of this form. While they may have a bulk interpretation \cite{Melton:2023lnz}, we do not supply an interpretation of such herein.} 
\begin{equation}
\label{eq:gluonop}
    \begin{split}
        \O^{+a,\varepsilon}_{\Delta}(z,\zb) &= \re^{-\im\varepsilon\pi(\Delta-1)/2}\;\lim_{b\to 0}\,N^+_{\Delta}\,J^a(z)\,V_{b(\Delta-1)/2}(z,\zb)\,, \\
        \O^{-a,\varepsilon}_{\Delta}(z,\zb) &= \re^{-\im\varepsilon\pi(\Delta+1)/2}\;\lim_{b\to 0}\, N^-_{\Delta}\, \bar J^a(z)\,V_{b(\Delta+1)/2}(z,\zb)\,,
    \end{split}
\end{equation}
where $\varepsilon=\pm 1$ labels whether the gluon is outgoing or ingoing and 
\begin{equation}
\begin{split}
    N^-_{\Delta} &= \lambda^{\Delta/2}\re^{\gamma_E} \sqrt{\re^{-2/b^2}\lambda^{-1/b^2}\pi b^3\sin(\pi/b^2)}\;\Gamma(\Delta+1)\,\re^{-\im\pi(\Delta+1)/2} \\
    N^+_{\Delta} &= \lambda^{(\Delta-1)/2}\,\Gamma(\Delta-1)\,\re^{-\im\pi(\Delta-1)/2}.
\end{split}
\end{equation}
Bulk amplitudes of  $\O^{\pm a,\varepsilon}_{\Delta}$ are then given by 2D CFT correlators of the operators on the right hand side. 
Taking the $b\to 0$ limit one finds the color-ordered correlator\footnote{Here we take  $\lim_{b\to 0}\frac{\sin(\pi/b^2)}{\sin\pi(1/b^2-\beta/2)}=\re^{\im\pi \beta/2}$, which follows if we take the limit with a small fixed phase for $b$. We note also that the celestial amplitudes are projected onto $\beta=0$ where this ratio is unity even before taking the limit. }
\begin{equation}
\begin{split}
   \Big\langle \O^{-,\varepsilon_1}_{\Delta_1}(z_1,\zb_1)\,\O^{-,\varepsilon_2}_{\Delta_2}(z_2,\zb_2)\prod_{j=3}^n\O^{+,\varepsilon_j}_{\Delta_j}(z_j,\zb_j)\Big\rangle &= \prod_j\re^{-\im\pi\varepsilon_j\hb_j}\Gamma(2\hb_j)\; \frac{z_{12}^4}{z_{12}\cdots z_{n1}}\; \cC_{2\hb_1,2\hb_2,\ldots,2\hb_n}\,.
\end{split}
\end{equation}
As shown in \cite{Melton:2023bjw}, this is precisely the expression for the MHV {\it leaf} amplitude obtained by integrating the gluon interaction over a single spacelike (but asymptotically null) $H^3$  slice in Minkowksi space. 

We conclude that the dressed Liouville theory supplied with the dictionary \eqref{eq:gluonop} correctly generates the Minkowskian  $H^3$ leaf amplitudes. 

%%%%%%%%%%%%%%%%%%%%%%%%%%%%%%%%
%%%%%%%%%%%%%%%%%%%%%%%%%%%%%%%%

\subsection{Euclidean leaves $\to$ Lorentzian leaves}

The full Minkowskian celestial amplitudes are given by integrals of the leaf amplitudes over all the leaves of a hyperbolic foliation of Minkowski space. These necessarily include integrals over the dS$_3$ leaves of the region spacelike separated from the origin as well as the $H^3$ leaves in the Milne region. Rather than compute dS$_3$ leaf amplitudes, we choose the simpler route of analytically continuing from Minkowski to Klein space, which  directly yields the Kleinian celestial amplitudes. The hyperbolic slices or leaves are then all Lorentzian $\mathrm{AdS}_3/\mathbb{Z}$ geometries and divide into two wedges 
containing points that are either timelike or spacelike separated from the origin. 

Analytically continuing to Klein space induces a corresponding continuation in the leaf amplitudes. In particular, the integrals over $H^3$ in the Witten diagram \eqref{lff} and its past-pointing counterpart become an integral over $\mathrm{AdS}_3/\mathbb{Z}$\ with an $\im\epsilon$ prescription in the bulk-to-boundary propagators.  This reproduces exactly the detailed formulae found in 
\cite{Melton:2023bjw} for the $n$-point MHV \emph{leaf} amplitudes. The leaf amplitudes now live on the boundary of Lorentzian AdS$_3/\Z$, known as the celestial torus.

%%%%%%%%%%%%%%%%%%%%%%%%%%%%%%%%
%%%%%%%%%%%%%%%%%%%%%%%%%%%%%%%%

\subsection{Lorentzian leaves $\to$ celestial amplitudes}

Finally, to get from leaf to celestial amplitudes, the holographic dictionary dictates that we extract the limit of the leaf amplitudes when the net conformal weights of the external particles obey
\be 
\beta \equiv \sum_j(\Delta_j-1) = 0\,.
\ee
This projection onto $\beta=0$ is performed by multiplying the leaf amplitudes with $\delta(\beta)$. As explained in \cite{Melton:2023bjw}, when this is combined with its image under exchanging the timelike and spacelike cycles on the celestial torus, and division by $8\pi^3$, one recovers exactly the $n$-gluon MHV celestial amplitudes.

%%%%%%%%%%%%%%%%%%%%%%%%%%%%%%%%
%%%%%%%%%%%%%%%%%%%%%%%%%%%%%%%%

\section*{Acknowledgements}

We are grateful to Eduardo Casali, Scott Collier, Laurent Freidel, Noah Miller, Nathan Seiberg and Tomasz Taylor for useful conversations. This work was supported by DOE grant de-sc/0007870, NSF GRFP grant DGE1745303, the Simons Collaboration on Celestial Holography and the  Gordon and Betty Moore Foundation and the John Templeton Foundation via the Black Hole Initiative.

%%%%%%%%%%%%%%%%%%%%%%%%%%%%%%%%
%%%%%%%%%%%%%%%%%%%%%%%%%%%%%%%%

\appendix

\section{Semiclassical Liouville Theory and Witten Diagrams}
\label{app:liouville}

We can take Liouville theory to have the action
\begin{equation}\label{SL}
    S_L = \frac{1}{4\pi}\int_{S^2} \d^2\xi\,\sqrt{\tilde{g}}\left(\tilde{g}^{ab}\,\partial_a\phi\,\partial_b\phi + Q\tilde{R}\phi + 4\pi \mu\,\re^{2b\phi}\right)\,.
\end{equation}
It is a 2D CFT with central charge
\begin{equation}
    c = 1+6Q^2 = 1+6\,(b+b^{-1})^2\,.
\end{equation}
In the action, $\tilde g$ denotes a fiducial metric on the 2-sphere, and $\tilde R$ is its scalar curvature. We will work in a coordinate patch on which $\tilde g$ is flat, at the cost of introducing boundary terms.

We are primarily interested in the semiclassical limit where $b \to 0$. In this limit, we work with a rescaled field $\phi_c = 2b\phi$, in terms of which the action becomes
\begin{equation}
    S_L = \frac{1}{4\pi b^2}\int_{D} \d^2z \left(\partial\phi_c\,\bpd\phi_c + 16\lambda\,\re^{\phi_c}\right) + \frac{1}{2\pi b^2}\oint_{\partial D}\phi_c\,\d\Theta + \frac{2}{b^2}\log R + \O(1)
\end{equation}
where $D=S^2-\infty$ is a disk of radius $R\to\infty$ that we equip with the flat metric $\d z\,\d\bz$. The combination $\lambda = \pi \mu b^2$ is held fixed as we take $b \to 0$. The equation of motion becomes 
\begin{equation}
    \partial\bpd\phi_c = 2\lambda\,\re^{\phi_c}
\end{equation}
Vertex operators of this theory take the form 
\begin{equation}
    V_\alpha(z,\zb) = \re^{2\alpha\phi(z,\zb)} = \re^{\alpha\phi_c(z,\zb)/b}\,.
\end{equation}
Their correlation functions can be defined through a path integral
\begin{equation}
    \Big\langle\prod_jV_{\alpha_j}(z_j,\zb_j)\Big\rangle = \int\mathcal{D}\phi\; \re^{-S_L}\prod_j \re^{\alpha_j\phi_c(z_j,\zb_j)/b}\,.
\end{equation}
Here,  $S_L$ scales as $1/b^2$, so \emph{heavy operators} with $\alpha = \eta/b$ will act as a source for the classical equations of motion. \emph{Light operators} with $\alpha$ scaling as $b$ will not change the classical saddle point, but instead just sample the value of a given classical saddle. 

Despite the absence of real classical saddles, the classical limit can be understood as a sum over complex saddles of the action \cite{Harlow:2011ny}. Extending the Liouville action to be a holomorphic function of a complex field $\phi$, the set of complex saddles is
\begin{equation}
    \phi_{c,N}^g = 2\pi\im\,\bigg(N+\frac12\bigg)-\log\lambda-2\log\big(|\alpha z+\beta|^2 + |\gamma z+\delta|^2\big)
\end{equation}
where $g = \begin{pmatrix} \alpha & \beta \\ \gamma & \delta \end{pmatrix}\in\SL(2,\C)$  and $N \in \mathbb{Z}$. 

In the semiclassical limit, we can write the all-light correlation function with weights $\al_j=b\sigma_j$ as an integral over the classical complex saddles:
\begin{equation}
    \Big\langle\prod_j V_{b\sigma_j}(z_j,\zb_j) \Big\rangle = \mp A(b)\sum_{N\in T} \int_{\SL(2,\mathbb{C})} \d g\; \re^{-S_L[\phi_{c,N}^g]}\prod_{j=1}^n\re^{\sigma_j\phi_{c,N}^g(z_j,\zb_j)}\,.
\end{equation}
Here $\d g = 4\,\delta^2(\alpha\delta-\beta\gamma-1)\,\d^2\alpha\,\d^2\beta \,\d^2\gamma\, \d^2\delta$ is the standard measure over SL(2,$\mathbb{C}$). The set indexing the sum is $T=\Z_{\geq0}$ and the sign is $-$ if $\Im(\beta-b^{-2}) > 0$; and $T=\Z_{<0}$ and the sign is $+$ if $\Im(\beta-b^{-2}) < 0$. Using the explicit form of the classical solutions, one obtains
\begin{multline}
         \Big\langle\prod_j V_{b\sigma_j}(z_j,\zb_j) \Big\rangle 
       =  \pm \lambda A(b)\sum_{N\in T}\re^{\frac{2+\log\lambda - 2\pi \im(N+1/2)}{b^2}}\,\lambda^{-\sum_j\sigma_j}\,\re^{2\pi \im(N+1/2)\sum_j\sigma_j} \\
       \times \int_{\SL(2,\C)} \d g\, \prod_{j=1}^n\frac{1}{(|\alpha z_j+\beta|^2 + |\gamma z_j+\delta|^2)^{2\sigma_j}}\,.
\end{multline}
At 3 points, this integral was performed explicitly, and the prefactor $A(b)$ is fixed by comparison with the DOZZ formula to be \cite{Harlow:2011ny}
\begin{equation}
    A(b) = \im\pi^{-3}b^{-3}\re^{-2\gamma_E}\,.
\end{equation}
Similar integrals for semiclassical light Liouville correlators have also been derived using fixed area techniques \cite{Seiberg:1990eb}. 

We now show that the integral over $\mathrm{SL}(2,\mathbb{C})$ becomes an $n$-point contact Witten diagram integrated over hyperbolic space $H^3$. Let $\lambda_j = (z_j,1)$. We can express the integral as
\begin{equation}
    \int_{\text{SL}(2,\mathbb{C})}\d g\, \prod_{j=1}^n\frac{1}{(|g\lambda_j|^2)^{2\s_j}}\,.
\end{equation}
Any element $g \in \text{SL}(2,\mathbb{C})$ admits the Iwasawa decomposition \cite{Freidel:2010xna} 
\begin{equation}
    g =  k\begin{pmatrix} \rho^{-1/2} & 0 \\ 0 & \rho^{1/2}\end{pmatrix} \begin{pmatrix} 1 & -y \\ 0 & 1 \end{pmatrix},\quad k \in \SU(2)\,,
\end{equation}
where $\rho\in(0,\infty)$ and $y\in\C$ get identified as coordinates on $H^3=\SL(2,\C)/\SU(2)$. In this parametrization, the factors in the integrand become Euclidean bulk-to-boundary propagators,
\begin{equation}
    \frac{1}{(|g\lambda|^2)^{2\sigma}} = \left(\frac{\rho}{\rho^2+|y-z|^2}\right)^{2\s}\,.
\end{equation}
Similarly, the SL(2,$\mathbb{C}$) measure decomposes as
\begin{equation}
    \d g = \d k\, \frac{\d\rho\, \d y\,\d\bar{y}}{\rho^3}
\end{equation}
where $\d k$ is the Haar measure on $\SU(2)$. Because the integrand is independent of the $\SU(2)$ element $k$, we can integrate over the $\SU(2)$ subgroup to obtain 
\begin{equation}
    \int_{\text{SL}(2,\mathbb{C})}\d g\, \prod_{j=1}^n\frac{1}{(|\alpha z_j+\beta|^2 + |\gamma z_j+\delta|^2)^{2\s_j}} = 2\pi^2\int_{H^3} \frac{\d\rho\, \d y\, \d\bar{y}}{\rho^3}\,\prod_{j=1}^n\left(\frac{\rho}{\rho^2+|y-z_j|^2}\right)^{2\s_j}\,.
\end{equation}
We finally recognize this as an $n$-point contact Witten diagram $\cC_{2\s_1,\ldots,2\s_2}$ integrated over $H^3$. Hence, we have, with $\beta = 2\,(\sum_j\s_j-2)$,
\begin{equation}
    \begin{split}
        \langle\prod_jV_{b\s_j}(z_j,\zb_j)\rangle &= \mp \frac{2\re^{-2\gamma_E}}{\im\pi b^3}\sum_{N\in T}\re^{\frac{2+\log\lambda - 2\pi \im(N+1/2)}{b^2}}\,\lambda^{1-\sum_j\sigma_j}\,\re^{2\pi \im(N+1/2)\sum_j\sigma_j}\,\cC_{2\s_1,\ldots,2\s_n} \\
        &= \frac{\re^{-2\gamma_E+2/b^2}\lambda^{1/b^2-1-\beta/2}}{\pi b^3}\csc\left(\pi\left(\frac{1}{b^2}-\frac{\beta}{2}\right)\right) \cC_{2\s_1,\ldots,2\s_n}\,.
        \end{split}
\end{equation}
\bibliographystyle{JHEP}
\bibliography{refs}

\end{document}